\documentclass{JASSS}

\usepackage{subcaption}
\usepackage{amssymb} 
\usepackage{bbm} 
\usepackage{url} 

\title{Mobility, Memory, and Network Structure in Agent-Based Models of Convention Tipping and Convergence}

\reviewcopy{false}

\author[1]{Joe Shymanski}
\author[1]{Garrick Springer}
\author[1]{Sandip Sen}
\affil[1]{University of Tulsa}

\email{sandip-sen@utulsa.edu}

\usepackage{natbib}
	\setcitestyle{authoryear,round,aysep={}}
	
\begin{document}
\maketitle 



\begin{abstract}
Tipping-point dynamics describe the critical conditions under which a committed minority drives a population to abandon an established convention in favor of a new one. We present a transparent agent-based model of this process, in which agents hold one of two behavioral states and a mobile committed minority attempts to overturn the incumbent convention. Our goal was to examine how localized mobility, bounded agent memory, and network topology jointly influence the tipping threshold. Using a custom agent-based simulation framework, we found that in many configurations, tipping becomes effectively inevitable: given sufficient time, the population always converges to the minority state. This observation motivated a complementary analysis focused on the pace of convergence rather than its feasibility. We introduce a unified predictive model that accurately estimates how structural and behavioral parameters determine the time required for complete adoption, showing that mobility is the dominant accelerator while memory and connectivity modulate convergence in systematic ways. Together, these results extend classical tipping-point research by linking structural and behavioral factors not only to the likelihood of convention change but also to the timescale on which it unfolds. While we frame the model in terms of convention-like binary behavioral adoption, the same mechanisms bear on norm change and other contagion-like social processes.
\end{abstract}


\parano{}



\section{Introduction}
\label{sec:introduction}

Social conventions and norms play a central role in shaping collective human behavior. Understanding how these shared behavioral regularities emerge, spread, and change is essential across sociology, economics, and artificial intelligence~\citep{andrighetto2022research,gelfand2024norm,morris19:norm,young1996economics}. Of particular interest are sudden shifts, or \textit{tipping points}, where a committed minority induces a majority to abandon an established convention in favor of a new one~\citep{andreoni2021predicting,centola2018experimental,de2018tipping,everall2023pareto,milkoreit2023social}. The size of this minority group, often called the \textit{critical mass}, has been studied in evolutionary game theory and through controlled experiments, with estimates of the threshold typically falling near 20--30\% of the population~\citep{centola2010spread,Centola2015,centola2018experimental,kandori1993learning,young1996economics}.

We use the term \textit{convention} deliberately. Following the norm-emergence tradition, we model a convention as a binary behavioral state that propagates through local reinforcement, rather than as a prescriptive rule backed by sanctions, values, or institutional authority. The mechanisms we study therefore apply most directly to convention-like norms and, more broadly, to any state that spreads through local interaction---opinions, innovations, or behaviors---with norm change being one important application rather than the full scope of the model. We make this modeling stance explicit because it scopes which social phenomena our results speak to and which they deliberately abstract away (see Related Work).

In practice, populations do not interact at random: individuals are connected through social networks that shape who influences whom and how quickly behaviors spread. Network topology therefore plays a central role in determining the conditions under which a new convention can take hold. In the context of multi-agent systems (MAS), such networked interactions offer insight into how collective conventions can be influenced or accelerated under different structural and behavioral conditions~\citep{baccino2023does,helfmann2021statistical,lopez2015modeling,morris19:norm}.

We focus on three factors---localized mobility, bounded memory, and network topology---because each is a consistently identified determinant of convention and norm spread in the emergence literature, yet they are rarely studied jointly. Mobility governs how committed actors are exposed to, and mix across, local neighborhoods~\citep{De2015,Masuda2017}; memory acts as behavioral inertia that reinforces the incumbent convention~\citep{Villatoro2009,centola2018experimental}; and topology fixes the reinforcement structure through which complex contagions either spread or stall~\citep{centola2010spread,watts1998collective}. These three dimensions are also largely orthogonal---one behavioral (memory), one dynamic (mobility), and one structural (topology)---which lets us isolate and characterize their individual and joint effects through systematic ablation rather than conflating them.

Our initial goal was to study how these three factors jointly affect tipping thresholds. Using a custom agent-based simulation framework, we examined how lattice and small-world~\citep{watts1998collective} networks, different neighborhood definitions~\citep{zaitsev2017generalized}, and mobile versus stationary agents influence the fraction of committed influencers required for large-scale adoption.

However, our experiments revealed that, within this dynamic setting, tipping appeared inevitable: given enough time, the minority state always spread to the full population. This observation motivated a complementary focus on the \textit{time required for convergence}. We therefore extended our analysis to quantify and model how structural and behavioral factors determine the pace at which convention change unfolds.

Prior work on tipping points typically focused on static networks or stateless agents~\citep{granovetter1978threshold,centola2018experimental}, limiting analysis to the minimal conditions required for change. By incorporating memory, localized movement, and structured interactions, our framework provides a richer setting that links traditional tipping-point theory with the temporal dynamics of convergence.

We emphasize that this is a \textit{theoretical}, mechanism-oriented agent-based model rather than an empirical predictor of any specific social system. Its purpose is to isolate and analytically characterize how four primitives---committed-minority seed fraction, mobility, memory, and connectivity---shape both whether and how quickly a convention tips, not to replicate the full cognitive or institutional richness of real norm dynamics. This stance follows the established practice in social simulation of using deliberately stylized models to expose mechanisms in a controlled setting, and it guides how we interpret our parameters and scope our claims throughout.

Our contributions are threefold:
\begin{itemize*}
    \item We introduce a transparent simulation framework for studying convention tipping under variations in topology, mobility, and memory.
    \item We show that in this dynamic system, tipping is assured in most contexts, allowing the analysis of convergence time as a complementary outcome of interest.
    \item We derive a unified predictive model that explains how structural and behavioral parameters jointly determine the timescale of adoption.
\end{itemize*}
This work builds on existing research on norm emergence, tipping points, and network science, but it extends prior approaches by showing how the same mechanisms that enable convention change also determine how quickly it unfolds.

The rest of this paper is organized as follows: we first review related work, then present our simulation process and modeling framework, and then report the results. The remaining sections discuss findings and conclusions and outline future directions.

\section{Related Work}
\label{sec:related-work}

The distinction between \textit{conventions} and \textit{social norms} is often blurred in the literature on collective behavior. In general, norms are viewed as conventions reinforced by social regulation or the threat of sanction, whereas conventions may persist through coordination alone~\citep{Mellema2020}. For the purposes of this study, we treat the two terms interchangeably, since our focus lies in the emergence and persistence of a minority behavior within a majority population. More broadly, the same mechanisms can describe any phenomenon that propagates through local interactions in a network, including opinions, ideologies, innovations, cultural preferences, diseases, or information cascades.

\subsection{Norm Emergence versus Normative MAS}
It is useful to locate our work within one of two strands of research on norms in multi-agent systems~\citep{morris19:norm}. The \textit{normative MAS} strand treats norms as explicit, prescriptive rules---often expressed with deontic concepts and enforced through sanctions, institutions, or designated authorities---and studies how agents reason about, adopt, or synthesize such rules. The \textit{norm-emergence} strand, by contrast, treats norms and conventions as implicit behavioral regularities that arise bottom-up from repeated local interaction and reinforcement, with a norm deemed to have ``emerged'' once a sufficient fraction of agents follow it. Our model belongs squarely to the second strand: agents hold a binary behavioral state and update it from local observation, with no explicit rules, sanctions, values, or hierarchy. This places our contribution alongside the tipping-point and convention-emergence literature rather than the deontic or institutional-norm literature, and it explains why norm-specific constructs such as sanction severity, goal feasibility, or influencer authority lie outside our present scope (we return to this in the Discussion).

\subsection{Positioning Relative to Critical-Mass and Convention-Change Models}
Interest in critical-mass and convention-change dynamics remains highly active. \citet{centola2018experimental} provided experimental evidence for tipping points in human social conventions, and \citet{andreoni2021predicting} showed in controlled experiments that the location of a tipping point can be predicted from measurable features of the population. On the modeling side, \citet{iacopini2022group} extended the committed-minority naming-game framework to higher-order group interactions, showing that the critical mass required to overturn a convention depends on the structure of group encounters---a structural sensitivity that parallels the topology and mobility effects we report. Most recently, \citet{ashery2025emergent} demonstrated that populations of large language model (LLM) agents spontaneously form shared conventions and exhibit committed-minority tipping, indicating that these dynamics extend even to AI-agent populations. Such LLM-based social simulations are an emerging but still unreliable methodology: recent assessments caution that current LLM agents exhibit inconsistent and biased behavior in social-simulation settings and advocate hybrid designs over wholesale replacement of mechanistic models~\citep{taillandier2025integrating,huang2024social}. This motivates our choice of a transparent, stylized agent-based model, in which every mechanism is interpretable and every parameter can be ablated, as a complementary and reproducible vantage point on the same critical-mass phenomena.

\subsection{Foundational Theories of Tipping Points}
Granovetter's threshold model~\citep{granovetter1978threshold} formalized how individuals adopt behaviors once a critical portion of their peers have already done so, explaining how small perturbations can trigger large-scale cascades. \citet{young1996economics} extended this reasoning to coordination games, showing that repeated local interactions can entrench or overturn conventions depending on early adopters.

Building on these theoretical foundations, Centola and colleagues~\citep{centola2010spread,Centola2015,centola2018experimental} combined empirical and simulation studies to demonstrate that tipping points for social conventions typically arise when 20--30\% of a population is committed to the new behavior. Their work emphasized that network clustering, memory of past interactions, and reinforcement from multiple neighbors are critical to successful convention change. \citet{gelfand2024norm} provide a comprehensive survey of this literature, showing that while tipping thresholds are a robust feature of collective dynamics, they are sensitive to structural and cognitive factors.

\subsection{Topology, Memory, and Minority Influence}
Network structure has been shown to strongly shape contagion and coordination processes. \citet{centola2010spread} found that clustered-lattice networks accelerate complex contagions compared to random graphs, as repeated reinforcement among neighbors facilitates behavioral adoption. \citet{Villatoro2009} explored related topology and memory effects, although their analysis centered on unbiased agents with multiple conventions rather than on committed minorities.

Subsequent work examined how the position and commitment of minority groups influence outcomes. \citet{baccino2023does} showed that small but resolute groups can drive majority adoption when embedded in favorable network locations, with influence amplified by the strength or visibility of commitment. These results reinforce Centola's insight that topology interacts with agent behavior to determine whether a minority can overturn a dominant convention.

\subsection{Influencer Characteristics and Mobility}
Most studies of convention diffusion assume fixed agent positions~\citep{helfmann2021statistical,Choi2010}, but mobility can fundamentally alter outcomes. Models of random walks~\citep{Masuda2017} and evolutionary dynamics with limited migration~\citep{De2015} suggest that movement enhances mixing, though typically in simplified or global forms. Our work extends this line of research by allowing influencers to move locally at each step, demonstrating that even modest mobility sharply lowers tipping thresholds and accelerates convergence.

\medskip
In summary, prior research establishes the importance of thresholds, reinforcement, and network structure in convention change. Our contribution unifies these insights by linking traditional tipping-point analyses with the temporal dynamics of convergence, showing how topology, memory, and mobility jointly determine both the emergence and the pace of adoption within a single simulation framework.

\section{Methodology}
\label{sec:methodology}

\subsection{Topological Generation}
We consider two network structures: square lattices and small-world networks, chosen for their relevance to social systems. Lattices capture localized, densely clustered interactions, while small-world networks introduce occasional long-range links~\citep{watts1998collective}. This structural diversity allows us to isolate the effects of connectivity and influencer mobility.

\subsubsection{Lattices.} We place $N=n^2$ agents on an $n \times n$ grid. Connectivity is determined by two Boolean parameters: \textit{torus}, which wraps edges to form a toroidal lattice, and \textit{Moore}, which distinguishes between von Neumann (4 neighbors) and Moore (8 neighbors) neighborhoods.

\subsubsection{Small-Worlds.} Small-world networks are generated using the Watts--Strogatz model~\citep{watts1998collective}. Each node is initially connected to $k$ nearest neighbors in a one-dimensional ring topology, and each edge is rewired with probability $p$. We fix $p=0.1$ to balance clustering with randomness, avoiding the fully regular ($p=0$) and fully random ($p=1$) extremes. Agents are assigned one node each in this ring, ensuring consistent spatial embedding across runs.

\subsection{Agent Types}
Agents are divided into two fixed types: \textit{ordinary} and \textit{influencer}. While type does not change during a simulation, each agent maintains a binary state representing its current convention, which can evolve through interaction.

\subsubsection{Ordinary.} Ordinary agents form the initial majority and all begin with state 0. They are \textit{stationary} within the network topology and update their state at each timestep by observing the states present in their neighborhood, both currently and over the previous $M$ timesteps stored in memory. Memory represents the number of past neighborhood observations retained by the agent.

We make this update rule explicit. Let $\mathcal{N}_i$ denote the network positions adjacent to agent $i$, and let $s_j(\tau)\in\{0,1\}$ be the state occupying position $j$ at time $\tau$ (occupancy can change as influencers move). At each step, agent $i$ tallies the states observed in its neighborhood over the current step and the previous $M$ steps,
\begin{equation}
    n_i^{(c)}(t) = \sum_{\tau=t-M}^{t}\ \sum_{j\in\mathcal{N}_i} \mathbbm{1}\!\left[s_j(\tau)=c\right], \qquad c\in\{0,1\},
    \label{eq:tally}
\end{equation}
and then adopts the locally dominant state, retaining its current state in the case of a tie:
\begin{equation}
    s_i(t+1) =
    \begin{cases}
        1 & \text{if } n_i^{(1)}(t) > n_i^{(0)}(t),\\[2pt]
        0 & \text{if } n_i^{(1)}(t) < n_i^{(0)}(t),\\[2pt]
        s_i(t) & \text{otherwise.}
    \end{cases}
    \label{eq:update}
\end{equation}
When $M=0$ the rule reduces to a memoryless majority over current neighbors. In this way, memory serves as temporal inertia, smoothing short-term fluctuations and reinforcing consistent local patterns.

The update in Equation~\ref{eq:update} is \textit{deterministic}: an agent always moves to its locally dominant state. A natural alternative is \textit{probabilistic} updating, in which an agent switches to state $1$ with probability proportional to $n_i^{(1)}(t)/\big(n_i^{(0)}(t)+n_i^{(1)}(t)\big)$, as in stochastic threshold and noisy-voter models. We adopt the deterministic rule because it makes the mechanisms maximally transparent and because, as shown in the Results section, it already drives the population to inevitable tipping given sufficient time. Probabilistic switching would introduce additional fluctuations that may shift the numerical values of thresholds and convergence times, but it would not change our qualitative findings: the same monotone dependence of tipping on seed fraction, mobility, and memory would persist, since stochastic updating preserves the direction of the local majority signal in expectation.

\subsubsection{Influencers.} A fraction $f$ of the population is designated as influencers, seeded with the alternative convention (state 1). Influencers are committed: they never change state, and they move with probability $m$ at each timestep. When moving, an influencer swaps positions with a randomly chosen ordinary neighbor.

We model mobility as a local swap, and restrict it to influencers, for three deliberate reasons. First, swapping conserves the degree sequence and overall structure of the network exactly, so that any change in outcomes can be attributed to \textit{movement} rather than to a concurrent change in topology---a confound that would arise under edge-rewiring forms of mobility. Second, restricting motion to the committed minority isolates the effect we care about, namely the exposure of a fixed set of seeds to new local neighborhoods, while keeping the majority substrate fixed as a controlled reference. Third, single-step local swaps match the small-step, local-diffusion assumption common to spatial mobility models, in which agents mix gradually with nearby regions rather than teleporting globally. This design ensures each node remains occupied and lets the minority convention diffuse gradually through the network. Allowing ordinary agents to move as well, or using non-local mobility models, is a natural extension that we leave to future work (see Future Work).

\subsection{Simulation Process}
Influencers were distributed evenly across the network using a Halton sequence with Owen scrambling to ensure well-dispersed yet distinct placements across runs~\citep{chi2005optimal}. For lattice networks, two-dimensional samples were rounded to the nearest grid coordinates; for small-world networks, one-dimensional samples were mapped directly to node IDs.

At each timestep, influencers moved with probability $m$, while ordinary agents updated their conventions based on their current neighborhoods and memory histories. The global adoption rate of the minority convention was recorded at every step. Simulations terminated upon reaching full adoption (ubiquity) or the maximum timestep limit.

\begin{figure}
    \centering
    \begin{subfigure}{.32\linewidth}
        \centering
        \includegraphics[width=\linewidth]{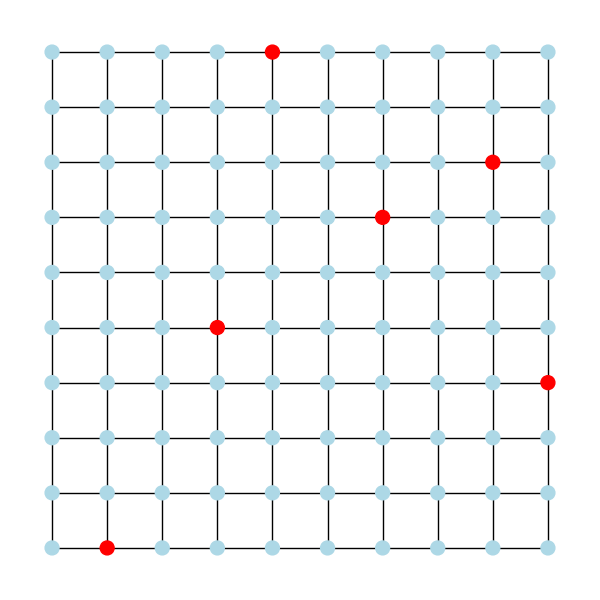}
        \caption{Step 0}
        \label{fig:lattice_step_0}
    \end{subfigure}
    \begin{subfigure}{.32\linewidth}
        \centering
        \includegraphics[width=\linewidth]{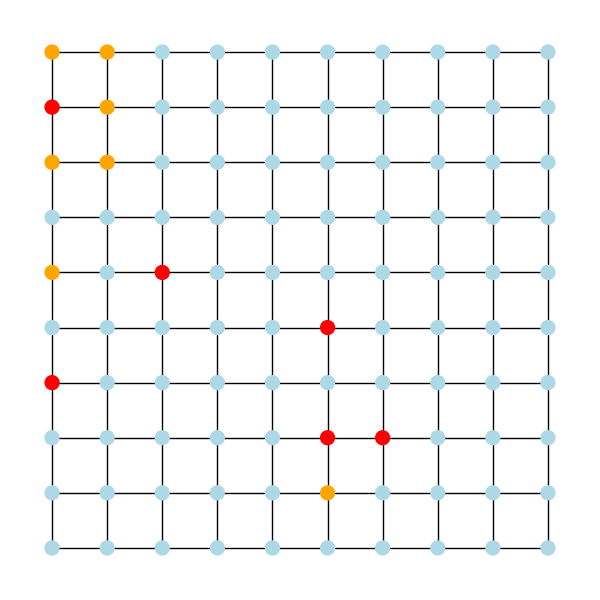}
        \caption{Step 500}
        \label{fig:lattice_step_500}
    \end{subfigure}
    \begin{subfigure}{.32\linewidth}
        \centering
        \includegraphics[width=\linewidth]{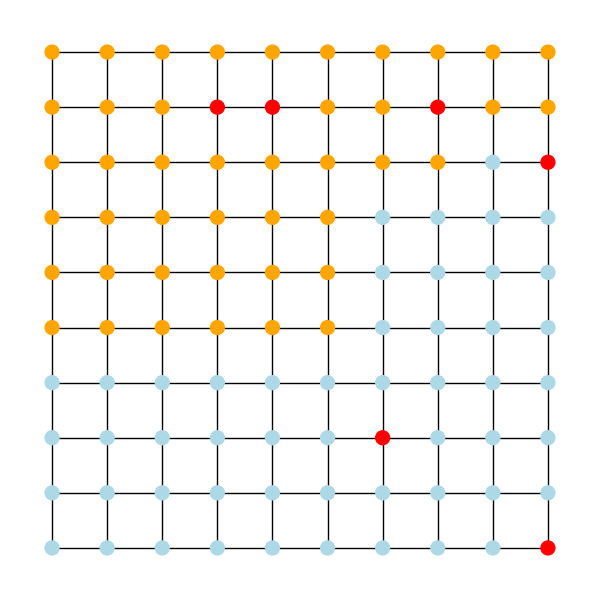}
        \caption{Step 1,000}
        \label{fig:lattice_step_1000}
    \end{subfigure}
    \begin{subfigure}{.32\linewidth}
        \centering
        \includegraphics[width=\linewidth]{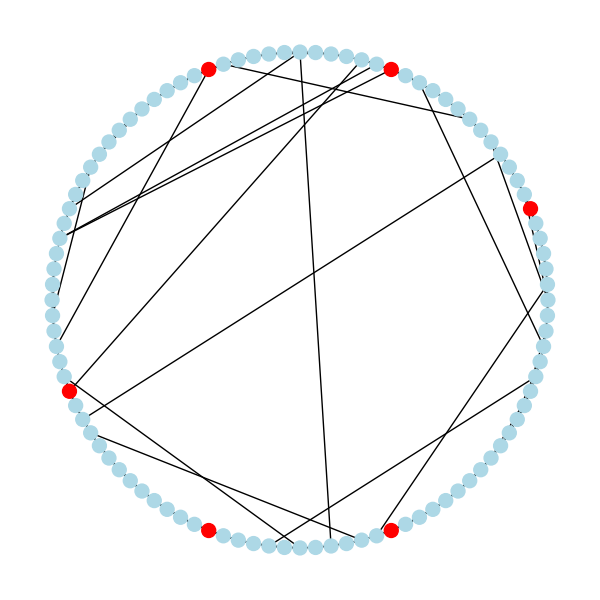}
        \caption{Step 0}
        \label{fig:smallworld_step_0}
    \end{subfigure}
    \begin{subfigure}{.32\linewidth}
        \centering
        \includegraphics[width=\linewidth]{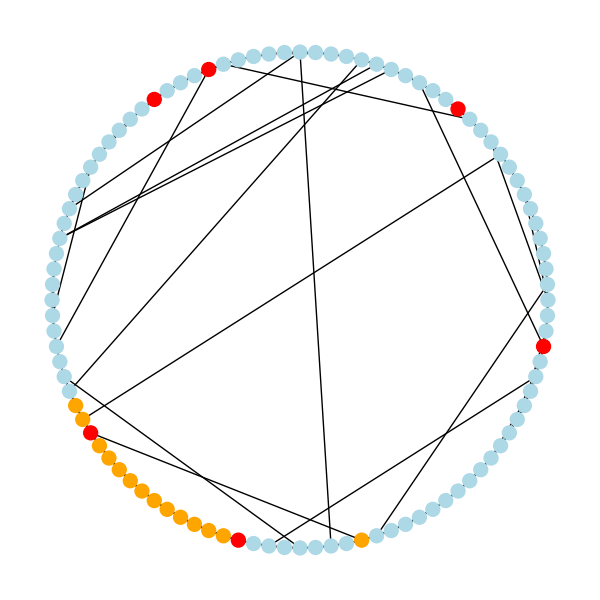}
        \caption{Step 500}
        \label{fig:smallworld_step_500}
    \end{subfigure}
    \begin{subfigure}{.32\linewidth}
        \centering
        \includegraphics[width=\linewidth]{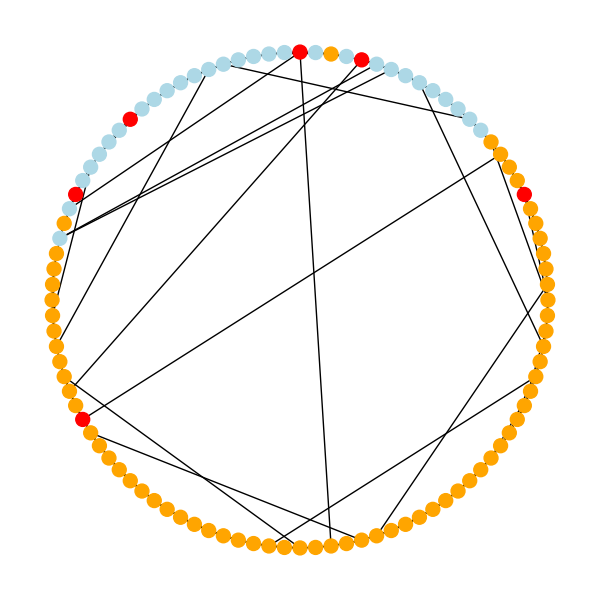}
        \caption{Step 1,000}
        \label{fig:smallworld_step_1000}
    \end{subfigure}
    \caption{Simulations using lattice (a--c) and small-world (d--f) networks. Blue nodes are ordinary agents in state 0, orange nodes are converted ordinary agents (state 1), and red nodes are influencers.}
    \label{fig:simulations}
\end{figure}

\begin{figure}
    \centering
    \begin{subfigure}{.49\linewidth}
        \centering
        \includegraphics[width=\linewidth]{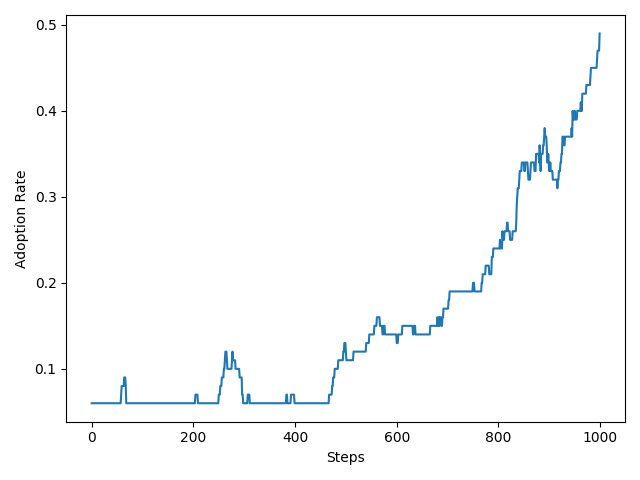}
        \caption{Lattice.}
        \label{fig:adoption_lattice}
    \end{subfigure}
    \begin{subfigure}{.49\linewidth}
        \centering
        \includegraphics[width=\linewidth]{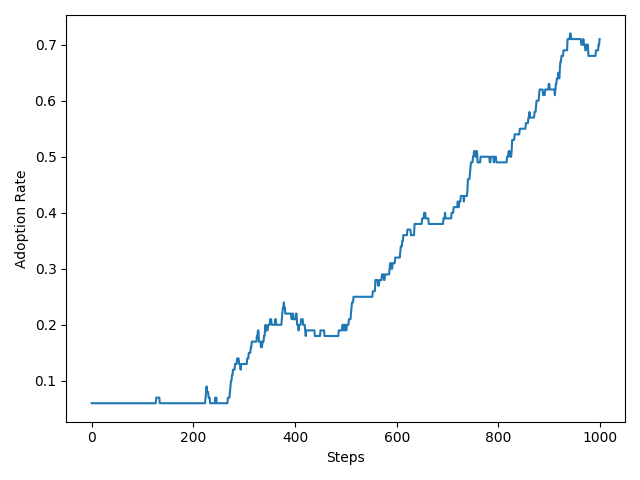}
        \caption{Small world.}
        \label{fig:adoption_smallworld}
    \end{subfigure}
    \caption{Adoption rates over time for the simulations shown in Figure~\ref{fig:simulations}.}
    \label{fig:adoption_rates}
\end{figure}

Figure~\ref{fig:simulations} illustrates representative lattice and small-world runs, with influencers shown in red, ordinary agents in state 0 in blue, and converted ordinary agents in orange. Figure~\ref{fig:adoption_rates} shows the corresponding adoption curves. Notably, adoption does not increase monotonically: as influencers move, local reversals occur, yet mobility ultimately drives convergence.

\subsection{Tipping Point Estimation}
\begin{figure}
    \centering
    \includegraphics[width=\linewidth]{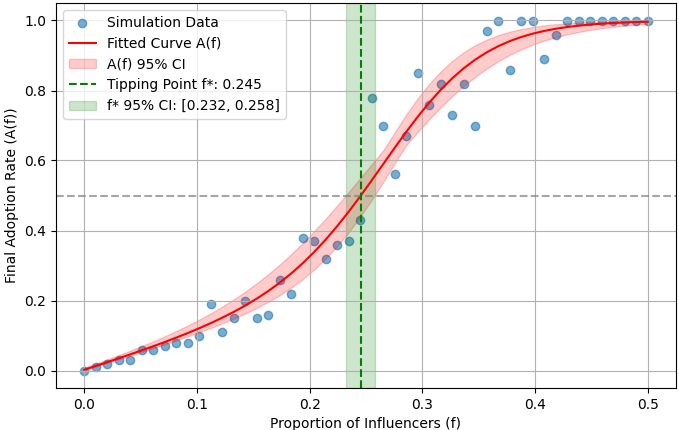}
    \caption{Example $A(f)$ curve fitted to simulation data for a single configuration (100-node lattice with von Neumann neighborhoods, no agent memory, and mobility rate of .01).}
    \label{fig:A_f_curve}
\end{figure}

To estimate the tipping threshold $f^*$ for each configuration, we ran 50 simulations at evenly spaced influencer fractions $f$, recording the final adoption rate $A$. We then fit these ($f$, $A$) samples to a logistic curve:
\begin{equation}
    A_{ord}(f) = \frac{1}{1+e^{-a(f - b)}},
    \label{eq:logistic}
\end{equation}
where $b$ is the midpoint and $a$ controls steepness. Because influencers never change state ($A_{inf}(f)=1$), the full adoption rate is given by:
\begin{equation}
    A(f) = fA_{inf}(f) + (1 - f)A_{ord}(f) = f + \frac{1 - f}{1 + e^{-a(f - b)}}
    \label{eq:objective}
\end{equation}
which incorporates both the linear contribution of influencers and the nonlinear response of ordinary agents. Parameters $a$ and $b$ were optimized using nonlinear least squares, and $f^*$ was obtained via root-finding at $A(f)=0.5$. Figure~\ref{fig:A_f_curve} illustrates a representative fit.

\begin{figure*}[t]
    \centering
    \includegraphics[width=\linewidth]{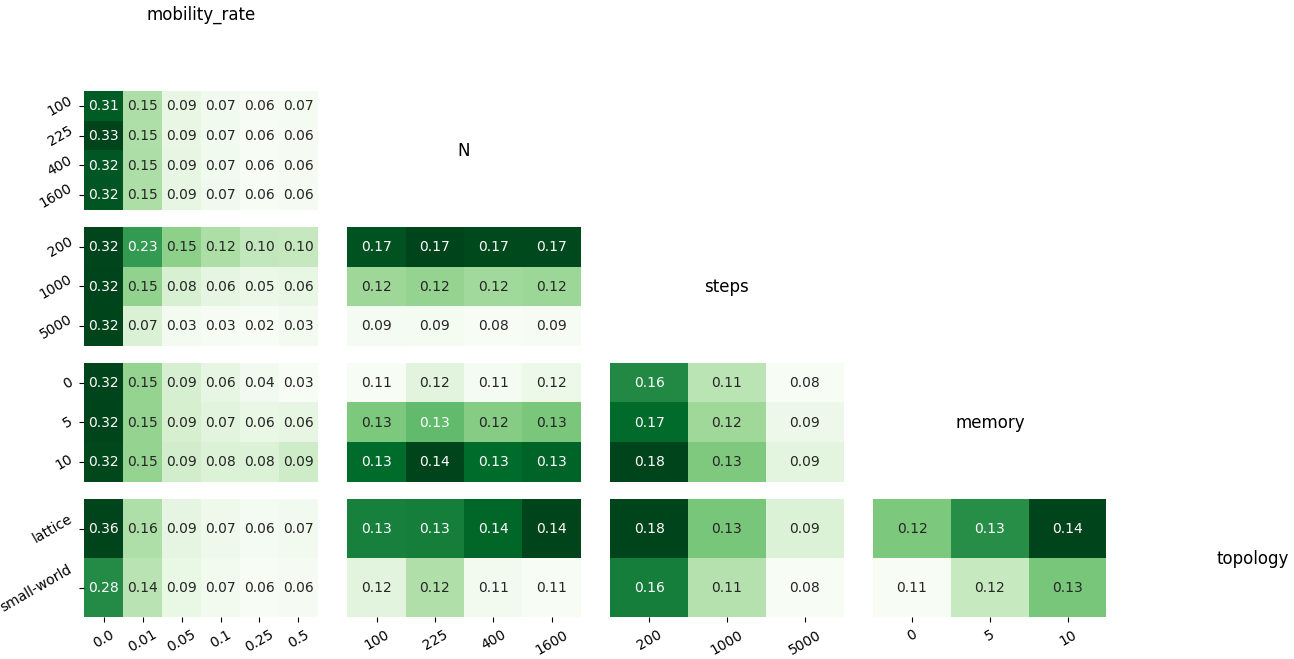}
    \caption{Pairwise tipping-point heatmaps over the five configuration parameters $m$, $N$, $steps$, $M$, and $topology$. The figure is a lower-triangular matrix of subplots: each subplot pairs two parameters, with one varying along the horizontal axis and the other along the vertical axis, and each cell shows the mean tipping threshold $f^*$ averaged over all configurations sharing that pair of values (marginalizing over the other three parameters). Axis tick labels give the discrete levels of each parameter. The diagonal is empty because a parameter is not plotted against itself, and the upper-right triangle is omitted because it would duplicate the lower-left subplots with axes transposed; the blank diagonal cell in the $topology$ row/column is this empty self-pairing, not missing data. Shading within each subplot is normalized independently to highlight relative trends, so colors are comparable within but not across subplots.}
    \label{fig:fstar_heatmaps}
\end{figure*}

\subsection{Time to Ubiquity Measure}
While tipping thresholds capture the conditions under which convention change becomes self-sustaining, they do not quantify how quickly convergence occurs once that process begins. To complement $f^*$, we define the \textit{time to ubiquity} (TTU), denoted $T$, as the first timestep at which the minority convention achieves full adoption. Let $A(t)\in[0,1]$ denote the global adoption rate---the fraction of agents in state $1$---at time $t$, and let $steps$ be the time horizon. Then
\begin{equation}
    T = \min\{\, t \in \{0,1,\dots,steps\} : A(t) = 1 \,\},
    \label{eq:ttu}
\end{equation}
and $T$ is left undefined (right-censored) when $A(t) < 1$ for all $t \le steps$, i.e.\ when ubiquity is not reached within the horizon. This measure captures the temporal dynamics of convention change, extending analysis beyond the onset of adoption to the speed of full convergence.

\section{Results}
\label{sec:results}

\subsection{Sensitivity Analysis}
We first conducted a systematic sensitivity analysis across five parameters: influencer mobility ($m$), agent memory ($M$), population size ($N$), maximum timesteps ($steps$), and network structure ($topology$). The 432 configurations are the full-factorial (Cartesian) product of the discrete levels of these five parameters:
\begin{equation*}
    \underbrace{6}_{m}\times\underbrace{3}_{M}\times\underbrace{3}_{steps}\times\underbrace{4}_{N}\times\underbrace{2}_{topology}=432,
\end{equation*}
with $m\in\{0,0.01,0.05,0.1,0.25,0.5\}$, $M\in\{0,5,10\}$, $steps\in\{200,1000,5000\}$, $N\in\{100,225,400,1600\}$, and $topology\in\{\text{lattice},\text{small-world}\}$ (using the von Neumann lattice and $k{=}4$ small-world as the canonical structure for each). For each configuration, 50 simulations were run at evenly spaced influencer fractions $f$, yielding $432\times 50 = 21{,}600$ $(f,A)$ samples. Each configuration was fitted to Equation~\ref{eq:objective} to extract the tipping threshold $f^*$. The full set of parameter levels and defaults is listed in Table~\ref{tab:variables} of the appendix.

Figure~\ref{fig:fstar_heatmaps} summarizes the pairwise effects among these parameters; each subplot shows mean $f^*$ for one pair of parameters, marginalizing over the other three (see caption for how to read the triangular layout). The clearest pattern is that $f^*$ declines steeply with increasing mobility, then plateaus at higher $m$. Longer interaction horizons ($steps$) also reduce the threshold, while greater memory capacity ($M$) inflates it. Lattice networks generally exhibit slightly higher tipping points than small-world networks, but both topology and population size show negligible influence compared to the other variables.

Spearman correlations confirm these trends: mobility ($\rho=-.69$), time horizon ($\rho=-.53$), and memory ($\rho=.15$) exhibit the strongest associations with $f^*$, whereas topology ($\rho=-.08$) and size ($|\rho|<.01$) are weak. These relationships indicate that the ease of convention change is governed primarily by the dynamic characteristics of movement, memory, and interaction frequency rather than by static network features.

\begin{figure}
    \centering
    \includegraphics[width=.5\linewidth]{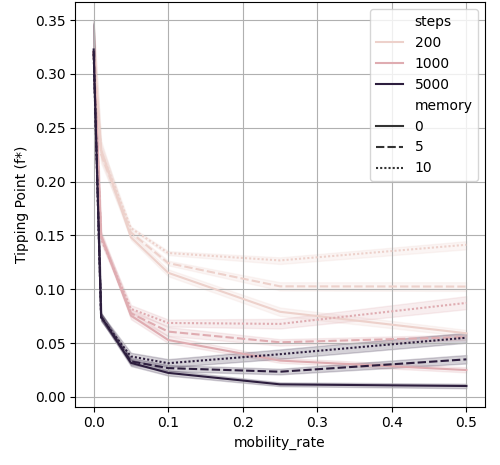}
    \caption{Trends in $f^*$ vs. mobility rate, memory, and time horizon ($steps$).}
    \label{fig:fstar_trendlines}
\end{figure}

Figure~\ref{fig:fstar_trendlines} illustrates the three most influential factors individually. $f^*$ decays rapidly with mobility but stabilizes at high $m$. Longer horizons decrease $f^*$, though this reduction can be offset by memory: as $M$ grows, agents resist change for longer, effectively flattening the mobility advantage. For example, with $m=0.5$, a 200-step memoryless simulation produces a similar threshold to one with 5,000 steps and $M=10$.

\subsection{Extended Analysis of Key Effects}
Beyond the overall sensitivity analysis, we conducted targeted experiments to more precisely characterize how memory, interaction horizon, and topology affect the tipping threshold $f^*$. Each experiment varied a single factor while holding others fixed at default values that can be found in Table~\ref{tab:variables} of Appendix~C.

These extended analyses confirmed the earlier trends while clarifying their functional forms. Increasing memory raised thresholds in a logarithmic fashion, consistent with the notion that recalling past interactions reinforces the incumbent convention but with diminishing returns. Conversely, extending the simulation horizon reduced $f^*$ according to a reciprocal decay, indicating that longer exposure provides more opportunities for influencers to trigger cascades. Both effects are stable across topologies.

Topology and mobility, however, exhibit a notable interaction. When influencers can move, denser lattices and small-world networks (via Moore neighborhoods or higher $k$) require larger influencer fractions to succeed, as each agent must counter more neighbors before switching. In contrast, when mobility is absent, thresholds are uniformly high across all topologies---reflecting that static influencers rarely achieve sufficient local clustering to propagate a new convention.

Overall, these extended tests reveal three dominant mechanisms: (1) memory acts as inertia, slowing convention change; (2) larger time horizons stimulate more opportunity for spread; and (3) topology amplifies or suppresses these forces depending on whether mobility is present. Fitted functional forms and detailed parameter tables are provided in Appendix~D. These relationships motivate the construction of a unified model in the next subsection that captures the speed of convergence.

\subsection{Time to Ubiquity Analysis}
We next analyzed the \emph{time to ubiquity} (TTU)---the number of simulation steps required for the influencer convention to achieve full adoption across the population. Each configuration was averaged over 50 trials on 225-node networks, systematically varying the influencer fraction ($f$), mobility rate ($m$), memory capacity ($M$), and mean node degree ($d$). These quantities were selected based on earlier sensitivity analyses that identified them as the dominant drivers of convention change.

\subsubsection{Modeling Objective.}
Rather than fitting a black-box predictor, we sought a compact, factorized \emph{equation} with correct limits and clear mechanism–term alignment. Each factor should correspond to a single driver (seed size, memory, mobility, connectivity) and reproduce the empirical shapes observed in the aggregated data.

\subsubsection{Empirical Patterns.} 
\begin{figure}[t]
    \centering
    \includegraphics[width=.5\linewidth]{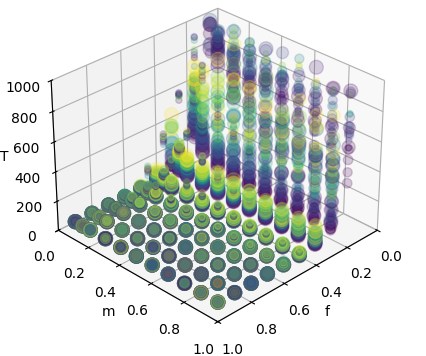}
    \caption{Mean time to ubiquity (TTU) across all configurations. Color encodes memory ($M$; low = purple, high = yellow), and point size encodes average node degree ($d$).}
    \label{fig:scatter_all}
\end{figure}

Figure~\ref{fig:scatter_all} visualizes the aggregated TTU data across all configurations. Each point represents an averaged run, with color encoding memory ($M$) and point size encoding connectivity ($d$). Convergence times decrease sharply with higher $f$ and $m$, increase with $M$, and vary nonlinearly with $d$: greater connectivity generally accelerates diffusion but can slow it when influencers are scarce and memory is large. These interactions highlight the need for a unified model that can reproduce all observed regimes without overfitting or losing interpretability.

\subsubsection{Model Structure and Rationale.} To capture these patterns, we derived a compact, factorized expression linking $T$ to the four primary variables while preserving their empirical shapes and mechanistic meanings.  
\begin{itemize*}
    \item The term $f^{-p_f}$ encodes how larger seed fractions ($f$) accelerate convergence following a power-law trend.
    \item The logarithmic factor $\ln(M+h_M)$ captures diminishing returns from memory, consistent with the extended analyses.
    \item The ratio $\frac{(m+g_m)^{p_m}}{(m+h_m)^{q_m}}$ models the steep early drop and plateau observed with increasing mobility ($m$).
    \item The connectivity term $\left(\frac{d}{d_{\min}}\right)^{e(f,M)}$ introduces a gated exponent $e(f,M)$ that flips sign only in the small-$f$, large-$M$ regime, reflecting when additional edges hinder rather than help adoption.
\end{itemize*}

Each component thus mirrors a distinct mechanism in the simulator---seeding, memory inertia, mobility diffusion, and network connectivity---making the model interpretable as well as predictive. Equation~\ref{eq:full} expresses these combined effects:
\begin{equation}
    T(f,M,m,d) = 1 + c f^{-p_f} \ln(M+h_M)
    \frac{(m+g_m)^{p_m}}{(m+h_m)^{q_m}}
    \left(\frac{d}{d_{\min}}\right)^{e(f,M)},
    \label{eq:full}
\end{equation}
where $d_{\min}=2-2/N$. The gated exponent is defined as:
\begin{equation}
    e(f,M)=e_{lo}+(e_{hi}-e_{lo})\,\sigma\!\left(\frac{f_c-f}{w_f}\right)\sigma\!\left(\frac{M-M_c}{w_M}\right),
    \quad \sigma(x)=\frac{1}{1+e^{-x}}.
    \label{eq:e_d}
\end{equation}

This factorization allows the equation to generalize across topologies while maintaining clear causal meaning. A detailed interpretation of each term and its empirical justification is provided below (see the subsection ``Interpreting the TTU Model'').

\subsection{Model Fitting and Evaluation}
The unified TTU equation was fitted with nonlinear least squares over more than 350{,}000 simulated runs spanning all tested values of $f$, $m$, $M$, and $d$. The fitted parameters (Table~\ref{tab:fitted-params}) yield strong fit quality, with $R^2=0.91$ and RMSE $\approx 57$, indicating that the model accounts for most of the variance and keeps typical errors modest relative to the scale of $T$.

Beyond raw fit, we also assess parsimony using Akaike’s Information Criterion (AIC), which penalizes unnecessary parameters. The unified model attains $AIC=52{,}144$, outperforming simpler baselines (Table~\ref{tab:model-comparison}). This shows that the additional terms improve predictive power sufficiently to justify their inclusion, rather than overfitting.

\begin{table*}[t]
    \centering
    \caption{Optimized parameters for Equation~\ref{eq:full}. The model achieves $R^2=0.91$, RMSE $=57.1$, and AIC $=52{,}144$.}
    \label{tab:fitted-params}
    \begin{tabular}{ccccccccccccc}
    \toprule
    $c$ & $p_f$ & $h_M$ & $g_m$ & $p_m$ & $h_m$ & $q_m$ & $e_{lo}$ & $e_{hi}$ & $f_c$ & $w_f$ & $M_c$ & $w_M$ \\
    \midrule
    7.39 & 1.52 & 5.33 & 0.074 & 19.11 & 0.069 & 18.95 & -2.44 & 9.31 & 0.32 & 0.022 & 148.14 & 108.64 \\
    \bottomrule
    \end{tabular}
\end{table*}

\subsubsection{Comparative Model Performance.}
We compare against three nested baselines that drop one mechanism at a time to test necessity and redundancy. Table~\ref{tab:model-comparison} reports $R^2$ and RMSE (goodness-of-fit) and AIC (parsimony with parameter penalty).

\begin{table}[t]
    \centering
    \caption{Model evaluation metrics for TTU prediction. Lower RMSE \& AIC and higher $R^2$ are better.}
    \label{tab:model-comparison}
    \begin{tabular}{lccc}
        \toprule
        \textit{Model} & $R^2$ & $RMSE$ & $AIC$ \\
        \midrule
        1. $cf^{-p_f}$ & 0.78 & 90.92 & 58,119.84 \\
        2. $cf^{-p_f}\ln\left(M+h_M\right)$ & 0.83 & 80.18 & 56,501.97 \\
        3. $cf^{-p_f}\ln\left(M+h_M\right)\frac{\left(m+g_m\right)^{p_m}}{\left(m+h_m\right)^{q_m}}$ & 0.85 & 76.14 & 55,842.96 \\
        \textbf{4. Equation~\ref{eq:full}} & \textbf{0.91} & \textbf{57.08} & \textbf{52,143.53} \\
        \bottomrule
    \end{tabular}
\end{table}

The incremental improvements from Models 1 $\rightarrow$ 4 reveal that each term is necessary to reproduce the observed dynamics. The largest $R^2$/RMSE gain appears when adding the mobility ratio, capturing the sharp reduction in $T$ as $m$ moves away from zero. The gated degree term delivers the final AIC improvement, removing residual bias in the small-$f$, large-$M$ corner without inflating complexity.

\subsubsection{Residual Analysis.} 
\begin{figure}[t]
    \centering
    \includegraphics[width=.5\linewidth]{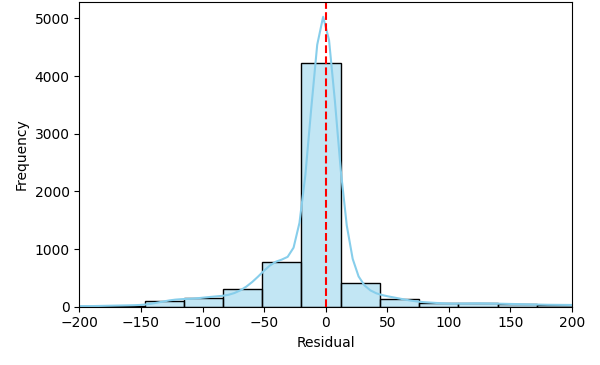}
    \caption{Histogram of residuals for $T(f,M,m,d)$. The near-symmetric shape indicates unbiased error distribution.}
    \label{fig:residual_hist}
\end{figure}

\begin{figure}[t]
\centering
\begin{subfigure}{.49\linewidth}
    \includegraphics[width=\linewidth]{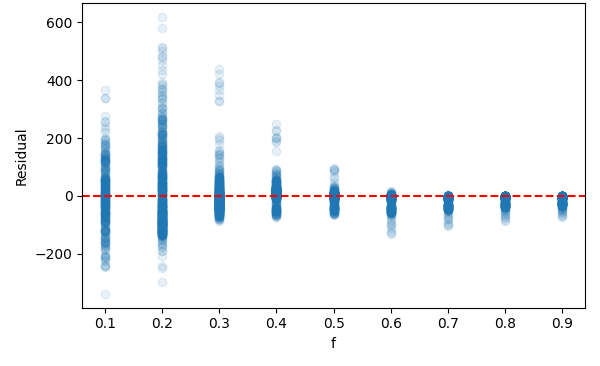}
    \caption{$f$}
    \label{fig:residual_f}
\end{subfigure}
\begin{subfigure}{.49\linewidth}
    \includegraphics[width=\linewidth]{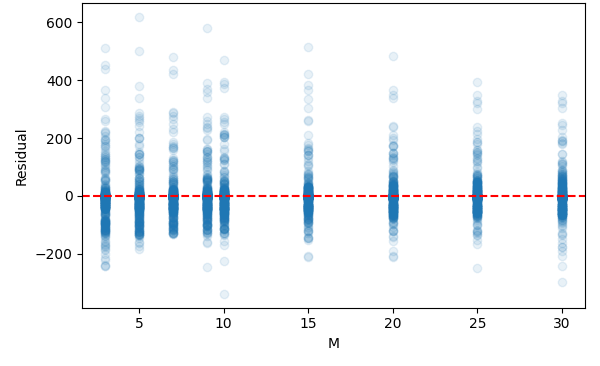}
    \caption{$M$}
    \label{fig:residual_M}
\end{subfigure}
\begin{subfigure}{.49\linewidth}
    \includegraphics[width=\linewidth]{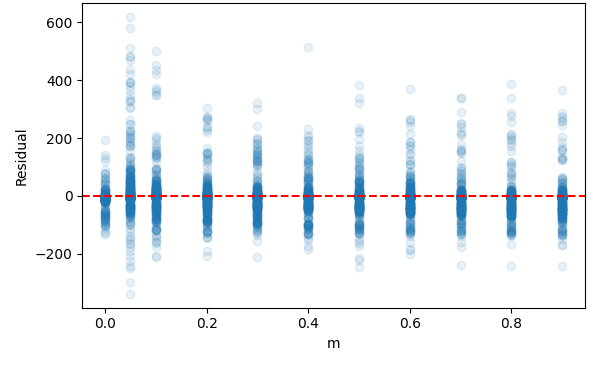}
    \caption{$m$}
    \label{fig:residual_m}
\end{subfigure}
\begin{subfigure}{.49\linewidth}
    \includegraphics[width=\linewidth]{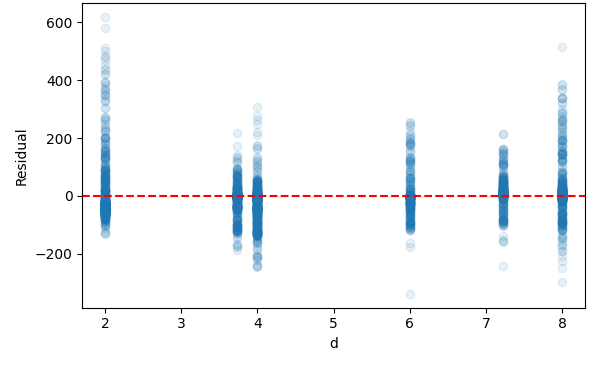}
    \caption{$d$}
    \label{fig:residual_d}
\end{subfigure}
\caption{Residuals of $T(f,M,m,d)$ plotted against each feature. The absence of systematic bias confirms model adequacy across the tested range.}
\label{fig:residual_feat}
\end{figure}

Figures~\ref{fig:residual_hist} and~\ref{fig:residual_feat} illustrate the model’s fit quality. Residuals are symmetric (Figure~\ref{fig:residual_hist}) and unbiased with respect to each feature (Figure~\ref{fig:residual_feat}), confirming that Equation~\ref{eq:full} generalizes across the tested domain.

The stepwise gains from Model 1 $\rightarrow$ Model 2 $\rightarrow$ Model 3 show that each mechanism contributes predictive power with minimal redundancy. The improvements are most pronounced in regimes with low $f$ or high $M$, where simpler models systematically underpredict convergence times. This demonstrates that the added complexity of Equation~\ref{eq:full} is necessary to capture the nonlinear interactions between influencer fraction, memory, mobility, and connectivity, while still yielding an interpretable functional form.

\subsection{Interpreting the TTU Model}
\label{sec:interpretation}
Equation~\ref{eq:full} is intentionally factorized so that each term corresponds directly to a mechanism in the simulator and satisfies the correct limiting behavior. This structure makes the model both explanatory and predictive, allowing each component to be read as a functional analogue of an underlying process governing convention convergence.

\subsubsection{Influencer Presence.} The multiplicative core $f^{-p_f}$ captures the universal ``more influencers $\Rightarrow$ faster convergence'' law and enforces the boundary $f \to 0^+ \Rightarrow T \to \infty$. The fitted exponent $p_f$ quantifies seed sensitivity: doubling $f$ reduces the expected TTU by approximately $2^{p_f}$. This term formalizes the observed power-law decay in convergence time with increasing influencer fraction.

\subsubsection{Memory.} The logarithmic factor $\ln(M + h_M)$ models the diminishing effect of memory on adoption speed. Larger $M$ increases inertia by reinforcing prior behaviors, but the influence of additional memory weakens at higher values, consistent with the extended analysis. The offset $h_M > 0$ ensures continuity at $M = 0$ and corresponds to the small baseline inertia observed even among memoryless agents.

\subsubsection{Mobility.} The rational mobility term $\frac{(m + g_m)^{p_m}}{(m + h_m)^{q_m}}$ produces the empirically observed ``elbow'' effect: as mobility increases from zero, $T$ drops sharply and then plateaus at higher $m$. This reproduces the finding that even small amounts of movement dramatically accelerate convention spread, after which additional movement yields diminishing returns. The small constants $g_m$ and $h_m$ position the knee near the range where mobility effects transition from marginal to saturated.

\subsubsection{Average Connectivity.} The gated connectivity term $\left(\frac{d}{d_{\min}}\right)^{e(f,M)}$ explains why degree can both help and hinder diffusion. The exponent $e(f,M)$ (Equation~\ref{eq:e_d}) flips sign only when influencers are scarce and memory is large, matching the empirical corner case where increased connectivity disperses influence rather than amplifying it. Outside this regime, higher degree consistently speeds convergence, capturing the typical ``more edges, faster diffusion'' relationship.

\subsubsection{Correct Limits and Parsimony.} The additive baseline $1 + \cdots$ anchors the model at one step for easy regimes (large $f$, moderate or high $m$) and ensures finite predictions as parameters vary. Together, these terms reproduce the correct asymptotic limits across all variables without cross-variable polynomials, preserving interpretability while maintaining high predictive accuracy.

\subsubsection{Synthesis.} Overall, each factor in Equation~\ref{eq:full} corresponds to an interpretable mechanism: seed size, memory inertia, mobility coupling, and degree-gated diffusion. Their combined behavior captures nuanced dynamics of convention convergence within a single predictive expression. This correspondence between mechanism and functional form grounds the model empirically and provides a transparent bridge to the theoretical discussion that follows.

\section{Discussion}
\label{sec:discussion}

Our results suggest that convention change in structured populations is shaped by two interdependent processes: first, reaching a tipping threshold $f^*$ that allows large-scale adoption to begin, and second, the ensuing convergence to full adoption over a time $T$. While analyzed separately, both reflect the same underlying mechanisms of mobility, memory, and network connectivity.

Although our model is deliberately abstract, each of its parameters maps onto an interpretable social quantity, which guards against the impression that the model's components are arbitrary. Table~\ref{tab:interpretation} summarizes these correspondences: the seed fraction $f$ is the committed minority, mobility $m$ is the rate of cross-neighborhood exposure or mixing, memory $M$ is the reinforcement history or behavioral inertia of ordinary agents, connectivity $d$ is the breadth of local exposure, and the time to ubiquity $T$ is the timescale over which an intervention would play out. We use these interpretations to frame the discussion that follows.

\begin{table}[t]
    \centering
    \caption{Mapping between model parameters and their social interpretations.}
    \label{tab:interpretation}
    \begin{tabular}{lll}
    \toprule
        \textit{Symbol} & \textit{Model quantity} & \textit{Social interpretation} \\
    \midrule
        $f$ & Influencer seed fraction & Committed minority / critical mass \\
        $m$ & Influencer mobility rate & Cross-community exposure and mixing \\
        $M$ & Agent memory length & Reinforcement history / behavioral inertia \\
        $d$ & Mean node degree & Breadth of local social exposure \\
        $T$ & Time to ubiquity & Convergence / intervention timescale \\
    \bottomrule
    \end{tabular}
\end{table}

The tipping thresholds we observed are consistently lower than those reported in prior empirical and theoretical studies~\citep{centola2018experimental,young1996economics}. This discrepancy arises from our local-interaction model: influencers interact repeatedly with neighbors rather than with randomly chosen partners, enabling minorities to reinforce one another even with uncoordinated movement. The effect of mobility is especially pronounced---without mobility, minorities require roughly 25\% of the population to succeed, but even minimal movement collapses this requirement by an order of magnitude.

Memory and time horizon also interact in predictable ways. Larger memory increases $f^*$, as agents become more resistant to change, but with diminishing returns consistent with earlier findings~\citep{centola2018experimental}. Longer horizons, by contrast, drive thresholds downward: given sufficient time, any network can eventually flip once the first conversion occurs. Degree effects are more nuanced—greater connectivity typically accelerates adoption, but under very small $f$ and high $M$, it can hinder spread by overwhelming agents with majority-state memories. Finally, we found no systematic effect of network size: thresholds and convergence times scale consistently, with $d_{\min} \to 2$ as $N \to \infty$, suggesting robustness to larger populations.

The unified TTU model integrates these insights, showing how the same structural and behavioral parameters that influence thresholds also determine convergence speed. Mobility again dominates: increasing $m$ not only lowers the required $f^*$ but also shortens $T$, whereas memory consistently slows adoption. The gated degree term explains why connectivity sometimes helps (when influencers are sufficient) and sometimes hurts (when they are scarce). Importantly, the model remains robust to network size, since $d_{\min}\!\to\!2$ as $N\!\to\!\infty$.

\subsection{Connection to Diffusion of Innovations}
The logistic form we fit to the adoption response (Equation~\ref{eq:logistic}) is not merely a convenient curve: it situates our results within the long-established theory of diffusion of innovations. \citet{rogers2003diffusion} described the cumulative adoption of an innovation across a population as an S-shaped (sigmoidal) curve, partitioning adopters into innovators, early adopters, an early and late majority, and laggards as the process passes through an initial slow phase, a rapid take-off once a critical mass is reached, and a final saturation. Our $A(f)$ curves reproduce exactly this shape: adoption is negligible below the tipping threshold $f^*$, rises steeply once the committed minority is large enough to trigger self-sustaining cascades, and saturates as the convention becomes ubiquitous. In this reading, the steepness parameter $a$ of the logistic captures how sharply the population transitions through Rogers' take-off phase, while the midpoint $b$ aligns with the critical mass at which the early majority begins to adopt. The same correspondence has long been used to model the diffusion of innovations on networks~\citep{Choi2010}, and it reinforces that our committed-minority mechanism is a structural account of the well-known S-curve rather than a new and unrelated phenomenon. The time to ubiquity $T$ then complements this static picture by quantifying \textit{how fast} the population traverses the S-curve once tipping is assured.

\subsection{Convention Switching and Churn}
Because ordinary agents update from a moving committed minority, adoption is not monotone: as influencers swap into and out of a neighborhood, locally converted agents can briefly revert before the convention takes hold permanently (visible as the dips in Figure~\ref{fig:adoption_rates}). The number of times agents switch state before convergence is therefore a meaningful secondary indicator of the ``churn'' or transient confusion a population experiences en route to consensus. Our present data record the global adoption trajectory $A(t)$ but not per-agent switch counts, so we report this only qualitatively: configurations that converge slowly---low $f$, low $m$, or high $M$---exhibit visibly more reversals along the way, consistent with the same inertia mechanisms that raise $T$. A direct measurement of switch counts as a function of mobility and memory is a concrete and inexpensive extension we have scoped for future work, and we expect it to track $T$ closely while sharpening the interpretation of memory as a stabilizer that suppresses churn.

\subsection{Value of Extensive Analysis}
The detailed parameter analyses were essential for constructing a grounded functional model rather than arbitrary curve fits. Each extended experiment directly informed one component of the unified TTU equation: the logarithmic dependence of $f^*$ on memory motivated the $\ln(M+h_M)$ term; the power-law decay of convergence time with increasing seed fraction inspired the $f^{-p_f}$ factor; the sharp early drop and plateau in mobility yielded the rational mobility term; and the mixed role of connectivity required the gated exponent $e(f,M)$. Together, these empirical trends shaped the model’s structure and ensured that each term corresponds to a distinct, interpretable mechanism. The breadth of experimentation thus served a methodological purpose—distilling the efficient, generalizable form of Equation~\ref{eq:full} across topologies—rather than simply expanding the dataset.

\subsection{Broader Implications}
The linked dynamics of tipping and convergence observed here mirror phenomena across social and biological systems. In online mobilization and influencer campaigns~\citep{bakshy2011everyone,centola2010spread,gonzalez2011dynamics}, modest increases in mobility (exposure to new micro-communities) can collapse both the required committed minority and the time to consensus, whereas long social ``memory'' (stable prior interactions or feeds) raises thresholds and slows convergence~\citep{flaxman2016filter,bicchieri2016norms}. In public health, the gated degree effect parallels contexts where high contact rates speed adoption of protective behaviors once seeding is adequate~\citep{centola2011experimental,bavel2020using}, but hinder change when initial uptake is scarce and prior behaviors are strongly reinforced~\citep{bicchieri2022nudging}. In civic settings, the framework clarifies when to invest in seeding (raising $f$), in mobility-enabling interventions (raising $m$), or in reducing memory-like reinforcement to accelerate convention change~\citep{centola2018experimental,young1996economics}.

\subsection{Limitations}
Our study has several limitations, the most fundamental of which is one of scope. By design, we model a convention as a binary behavioral state that spreads through local reinforcement; we therefore deliberately abstract away the features that distinguish full \textit{social norms} from bare conventions. In particular, our agents have no values or preferences that would make some conventions more desirable than others, no sanctions or punishment for non-conformity, no notion of norm legitimacy or institutional backing, no expectations about what others believe one ought to do, and no hierarchy or differential authority among influencers. These are precisely the norm-specific constructs studied in the normative-MAS and institutional-norm literatures, and their omission means our results speak to convention-like adoption rather than to the full richness of normative change. We see this not as a flaw but as a controlled boundary: the present framework isolates the structural and behavioral mechanisms (mobility, memory, connectivity, seeding) cleanly, and sanction-, value-, or hierarchy-based dynamics could be layered atop it as explicit extensions.

Beyond scope, several modeling choices bound generality. First, simulations were restricted to relatively small, homogeneous populations; although $f^*$ showed little sensitivity to network size, larger or more heterogeneous networks may exhibit new effects. Second, only the committed minority moves, and it does so by local swaps; allowing ordinary agents to move, or using non-local mobility, may change the quantitative picture. Third, influencers were modeled as stubborn but unstrategic; real actors often adapt or coordinate, potentially lowering thresholds further. Fourth, we focused on binary adoption, leaving open how multiple competing conventions or weighted preferences interact with tipping and ubiquity. Finally, the update rule is deterministic; stochastic switching would add realism at the cost of transparency. These extensions present natural directions for future work.

\section{Conclusion}
\label{sec:conclusion}

This work presents a unified perspective on convention change by linking the conditions that enable large-scale adoption with the dynamics that govern its completion. Using a custom agent-based simulation framework, we show that:
\begin{itemize*}
    \item Even modest mobility drastically lowers tipping thresholds and accelerates convergence.
    \item Memory increases resistance to change but with diminishing returns, elevating the tipping threshold $f^*$ and lengthening convergence time $T$.
    \item Connectivity generally facilitates diffusion, though under sparse seeding and high memory it can impede adoption.
\end{itemize*}

These findings reveal the surprising efficiency of uncoordinated minorities: without communication or strategy, a small mobile faction can overturn entrenched conventions in networks of hundreds or thousands. Beyond social conventions and norms, this framework extends naturally to opinion dynamics, ideological diffusion, political mobilization, and epidemic modeling, complementing classical contagion approaches such as SIR~\citep{weiss2013sir}.

Crucially, the extensive analyses throughout this study were not exploratory for their own sake—they identified the functional dependencies of mobility, memory, and connectivity that underpin both tipping and convergence. These empirical insights directly shaped the unified TTU equation, ensuring that each term reflects a measurable relationship rather than a curve-fitting artifact. By grounding the model in systematic simulations and theoretical limits, we provide an interpretable and general framework for predicting not only \emph{when} convention change occurs, but also \emph{how quickly} it completes.

To ensure reproducibility, the full simulation framework and analysis scripts are archived in the CoMSES Computational Model Library, with documentation following the ODD protocol summarized in Appendix~A. The model can be found at\\
\url{https://www.comses.net/codebase-release/4d4ad689-1d32-48a0-93a4-594bc78a0eea/}.

\section{Future Work}
\label{sec:future-work}

Several promising extensions remain for future exploration:
\begin{itemize*}
    \item \textbf{Entrenched agents.} Real populations include individuals with strong inertia or vested interests. Modeling agents who resist switching back after adoption could reveal hysteresis effects, where reversal thresholds differ from initial tipping points.
    \item \textbf{Punishment and enforcement.} Incorporating sanctions against norm violators may clarify how deterrence mechanisms interact with committed minorities and influence stability.
    \item \textbf{Influencer persistence.} Minority influencers may lose motivation or energy over time, reducing mobility or reverting to the majority convention; modeling this decay could yield richer temporal dynamics.
    \item \textbf{Strategic and generalized mobility.} Future work could examine coordinated or adaptive movement policies, where influencers adjust their rate or direction of travel based on local feedback or learned trajectories. A complementary direction is to relax our modeling choice that only the committed minority moves, allowing ordinary agents to relocate as well, and to compare local swaps against non-local mobility models.
    \item \textbf{Quantifying switching and churn.} Recording the number of times each agent changes state before convergence would measure the transient churn a population undergoes, sharpening the interpretation of memory as a stabilizer and providing a second outcome to validate against the time to ubiquity.
    \item \textbf{Seeding strategies.} Investigating targeted placements or adaptive seeding methods may identify more efficient pathways to convention change than uniform initialization.
    \item \textbf{Network diversity.} Extending the analysis to scale-free, hub-and-spoke, or multilayer networks could generalize the findings to heterogeneous real-world systems.
    \item \textbf{Prediction and intervention.} Developing early-warning indicators from adoption curves could enable active control of tipping dynamics, such as adding influencers or boosting mobility to accelerate desired outcomes.
    \item \textbf{Multiple conventions and competition.} Allowing concurrent conventions to spread may reveal coexistence equilibria, oscillations, or cyclic dominance akin to ecological models.
    \item \textbf{Heterogeneous agents.} Introducing variability in susceptibility, memory, or mobility could uncover mixed strategies where some agents act as early adopters while others remain long-term holdouts.
    \item \textbf{Empirical validation.} Applying the framework to real-world datasets—such as social media cascades, organizational change, or epidemic diffusion—would test external validity and illuminate which mechanisms dominate in practice.
\end{itemize*}

\endparano



% ...

\section{Appendix A: ODD Model Description}
\label{sec:odd}

For transparency and reproducibility, we summarize the model using the Overview, Design concepts, and Details (ODD) protocol. Notation follows the main text, and the parameter levels and defaults referenced below are listed in Table~\ref{tab:variables}.

\subsection{Overview}

\subsubsection{Purpose.} The model is a theoretical, mechanism-oriented agent-based model whose purpose is to isolate how four primitives---the committed-minority seed fraction ($f$), influencer mobility ($m$), agent memory ($M$), and network connectivity ($d$)---jointly determine whether and how quickly a population abandons an incumbent binary convention in favor of one promoted by a committed minority. It is not intended as an empirically calibrated predictor of any specific social system.

\subsubsection{Entities, state variables, and scales.} The model contains a single agent type partitioned into two roles. \textit{Ordinary} agents (the initial majority) carry a mutable binary state $s_i\in\{0,1\}$ and are fixed to a network position. \textit{Influencer} agents (the committed minority, a fraction $f$ of the population) are permanently in state $1$ and may relocate. Agents occupy the nodes of a static network---either a square lattice (with optional torus wrapping and von Neumann or Moore neighborhoods) or a Watts--Strogatz small-world graph (degree $k$, rewiring probability $p$). Each ordinary agent additionally stores the states observed in its neighborhood over the previous $M$ timesteps. One timestep corresponds to one synchronous update of all agents; a run lasts at most $steps$ timesteps.

\subsubsection{Process overview and scheduling.} At each timestep: (1) every influencer relocates with probability $m$ by swapping positions with a randomly chosen ordinary neighbor; (2) every ordinary agent updates its state synchronously via the local majority-over-memory rule (Equations~\ref{eq:tally}--\ref{eq:update}); (3) the global adoption rate $A(t)$ is recorded. The run terminates when $A(t)=1$ (ubiquity) or when $steps$ is reached.

\subsection{Design Concepts}
\begin{itemize*}
    \item \textbf{Basic principle.} Convention change is modeled as a complex contagion driven by local reinforcement from a committed minority, in the norm-emergence tradition.
    \item \textbf{Emergence.} The tipping threshold $f^*$ and time to ubiquity $T$ are emergent outcomes of local interactions, not imposed.
    \item \textbf{Sensing and interaction.} Ordinary agents sense only the states present in their network neighborhood, aggregated over their memory window; there is no global information or direct communication.
    \item \textbf{Stochasticity.} Randomness enters through influencer placement (a scrambled Halton sequence) and through probabilistic movement at rate $m$. State updates themselves are deterministic.
    \item \textbf{Observation.} The adoption trajectory $A(t)$ is observed every step; from it we derive $f^*$ (via the logistic fit of Equation~\ref{eq:objective}) and $T$ (Equation~\ref{eq:ttu}).
\end{itemize*}

\subsection{Details}

\subsubsection{Initialization.} The population is placed on the chosen network. A fraction $f$ of nodes is designated as influencers using a scrambled Halton sequence (Appendix~B) and set to state $1$; all remaining (ordinary) agents start in state $0$ with empty memory.

\subsubsection{Input data.} The model uses no external input data; all dynamics are generated endogenously.

\subsubsection{Submodels.} The movement submodel is the local swap described in the Methodology section. The update submodel is the majority-over-memory rule of Equations~\ref{eq:tally}--\ref{eq:update}. The measurement submodels are the logistic tipping-point estimator (Equations~\ref{eq:logistic}--\ref{eq:objective}) and the time-to-ubiquity measure (Equation~\ref{eq:ttu}).

\section{Appendix B: Influencer Placement}
\label{sec:halton}

Influencers were placed using a Halton sequence with Owen scrambling to ensure even yet varied distributions across runs. For lattice networks, two-dimensional Halton samples were generated, rounded to the nearest integer, and mapped to grid coordinates. For small-world networks, one-dimensional samples were directly mapped to node IDs. This approach yields balanced but stochastic seeding, reducing sampling bias while avoiding artificial clustering. The same randomized seed configuration was applied consistently across both lattice and small-world topologies to ensure comparability.

\section{Appendix C: Simulation Parameters}
\label{sec:params}

Table~\ref{tab:variables} lists the variables used in the simulations, their admissible values, and defaults when not explicitly varied. Dynamic parameters such as \(m\), \(M\), and \(f\) drive social processes, while structural parameters (e.g., \(k\), \(p\)) define the interaction graph. Holding most constants at defaults enables clean ablation of key mechanisms in sensitivity studies.

\begin{table}
    \centering
    \caption{Variable descriptions, possible values, and default values.}
    \label{tab:variables}
    \begin{tabular}{llll}
    \toprule
        \textit{Variable} & \textit{Description} & \textit{Possible Values} & \textit{Default} \\
    \midrule
        $A$ & Final adoption rate & $[0,1]$ & -- \\
        $dist$ & Initial influencer distribution & \{even\} & even \\
        $f$ & Fraction of influencers & $[0,1]$ & -- \\
        $f^*$ & Tipping threshold & $[0,1]$ & -- \\
        $k$ & Small-world: avg.\ node degree & $\mathbb{N}$ & 4 \\
        $M$ & Agent memory (steps) & $\mathbb{N}$ & 5 \\
        $m$ & Influencer mobility rate & $[0,1]$ & 0.05 \\
        $Moore$ & Lattice: Moore neighborhoods? & \{true, false\} & false \\
        $N$ & Number of agents/nodes & $\mathbb{N}$ & 100 \\
        $p$ & Small-world: rewiring prob. & $[0,1]$ & 0.1 \\
        $steps$ & Time horizon (max steps) & $\mathbb{N}$ & 200 \\
        $topology$ & Lattice or small-world & \{lattice, small-world\} & -- \\
        $torus$ & Lattice: torus wrapping? & \{true, false\} & false \\
    \bottomrule
    \end{tabular}
\end{table}

\section{Appendix D: Extended Analysis of Key Effects}
\label{sec:extended}

These focused tests isolate single drivers while holding others at default settings, clarifying functional forms that later inform the unified TTU equation (Equation~\ref{eq:full}) in the main text.

\subsection{Memory}
Isolating memory revealed a logarithmic dependence of the tipping threshold:
\begin{equation}
    f^*(M) = a\,\ln(M + b).
    \label{eq:memory_supp}
\end{equation}
The fit (Figure~\ref{fig:fstar_extended_supp}) achieved $R^2 \approx 0.97$. Although the function lacks a horizontal asymptote, $f^*$ cannot exceed $0.5$; unrealistically large $M$ would be required to approach that bound.

\subsection{Steps}
\begin{figure}
\centering
\begin{subfigure}{.49\linewidth}
    \centering
    \includegraphics[width=\linewidth]{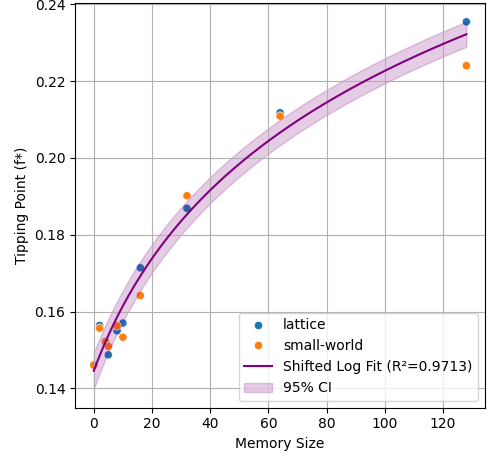}
    \caption{$M$}
    \label{fig:fstar_memory_supp}
\end{subfigure}
\begin{subfigure}{.49\linewidth}
    \centering
    \includegraphics[width=\linewidth]{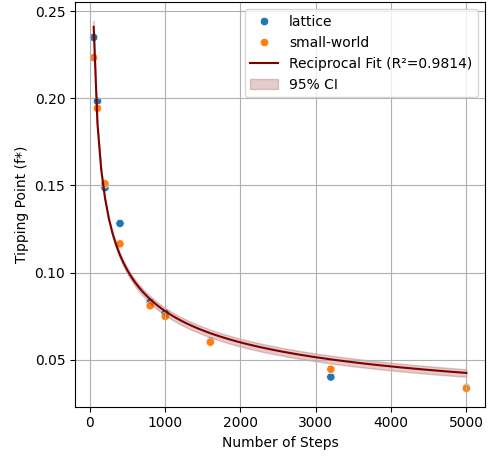}
    \caption{$steps$}
    \label{fig:fstar_steps_supp}
\end{subfigure}
\caption{Extended analysis of $M$ and $steps$, showing supplementary $f^*$ samples with other parameters fixed to defaults.}
\label{fig:fstar_extended_supp}
\end{figure}

Varying the maximum number of timesteps while fixing other parameters yielded a reciprocal decay:
\begin{equation}
    f^*(steps) = \frac{a}{steps^b},
    \label{eq:steps_supp}
\end{equation}
with $R^2 \approx 0.98$ (Figure~\ref{fig:fstar_extended_supp}). This implies thresholds can be driven arbitrarily close to zero given sufficient time.

\subsection{Topology}
Table~\ref{tab:topology_supp} reports the effects of lattice parameters (torus and Moore neighborhoods) and small-world degree \(k\) across mobility rates \(m\). With mobile influencers, denser networks require larger \(f\) to succeed; with immobile influencers, thresholds remain high across all structures.

\begin{table}
    \centering
    \caption{Topology and mobility effects on $f^*$ in lattices (left) and small-worlds (right); other parameters fixed.}
    \label{tab:topology_supp}
    \begin{tabular}{ccccc}
    \toprule
        $m$ & $torus$ & $Moore$ & $f^*$ & 95\% CI \\
    \midrule
        0 &   &   & .354 & [.322, .381] \\
        0 & \checkmark &   & .399 & [.363, .428] \\
        0 &   & \checkmark & .317 & [.293, .335] \\
        0 & \checkmark & \checkmark & .338 & [.307, .358] \\
    \midrule
        .10 &   &   & .121 & [.113, .127] \\
        .10 & \checkmark &   & .134 & [.129, .139] \\
        .10 &   & \checkmark & .152 & [.145, .158] \\
        .10 & \checkmark & \checkmark & .178 & [.162, .189] \\
    \midrule
        .25 &   &   & .110 & [.105, .116] \\
        .25 & \checkmark &   & .127 & [.124, .129] \\
        .25 &   & \checkmark & .151 & [.138, .162] \\
        .25 & \checkmark & \checkmark & .194 & [.180, .203] \\
    \bottomrule
    \end{tabular}
    \hspace{1em}
    \begin{tabular}{cccc}
    \toprule
        $m$ & $k$ & $f^*$ & 95\% CI \\
    \midrule
        0 & 2 & .379 & [.364, .394] \\
        0 & 4 & .280 & [.268, .291] \\
        0 & 6 & .347 & [.333, .350] \\
        0 & 8 & .319 & [.307, .331] \\
    \midrule
        .10 & 2 & .121 & [.114, .129] \\
        .10 & 4 & .122 & [.117, .127] \\
        .10 & 6 & .144 & [.141, .147] \\
        .10 & 8 & .164 & [.161, .166] \\
    \midrule
        .25 & 2 & .084 & [.079, .089] \\
        .25 & 4 & .098 & [.093, .102] \\
        .25 & 6 & .127 & [.125, .130] \\
        .25 & 8 & .168 & [.167, .168] \\
    \bottomrule
    \end{tabular}
\end{table}

In static environments, dense connections offer little advantage because influence cannot migrate beyond fixed neighborhoods. Under mobility, the same connectivity becomes an accelerant: influencers traverse boundaries and bridge clusters, explaining the nonlinear thresholds seen in the broader analyses.

\section{Appendix E: Extended Time-to-Ubiquity Analysis}
\label{sec:extended-ttu}

Here we expand on time-to-ubiquity (TTU) patterns across topologies. While the unified model (Equation~\ref{eq:full}) provides a compact predictor, visualizing raw data clarifies structural differences between grids and small-worlds. Each point aggregates 50 trials per configuration, smoothing temporal noise and highlighting stable macro-level trends.

\subsection{Scatterplots by Topology}

\begin{figure}
\centering
\begin{subfigure}{.49\linewidth}
    \includegraphics[width=\linewidth]{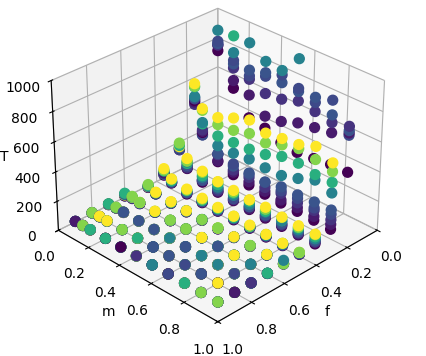}
    \caption{Lattice, von Neumann}
    \label{fig:scatter_lat_vn}
\end{subfigure}
\begin{subfigure}{.49\linewidth}
    \includegraphics[width=\linewidth]{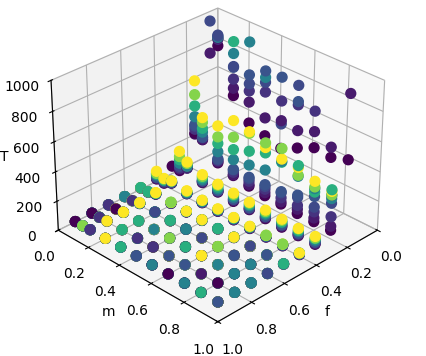}
    \caption{Torus, von Neumann}
    \label{fig:scatter_tor_vn}
\end{subfigure}
\begin{subfigure}{.49\linewidth}
    \includegraphics[width=\linewidth]{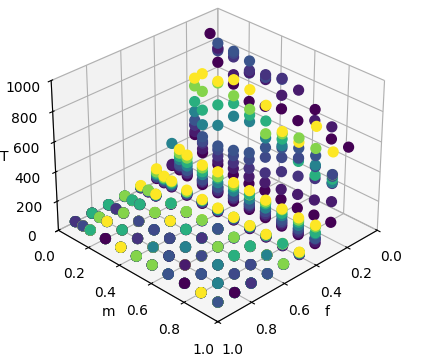}
    \caption{Lattice, Moore}
    \label{fig:scatter_lat_mo}
\end{subfigure}
\begin{subfigure}{.49\linewidth}
    \includegraphics[width=\linewidth]{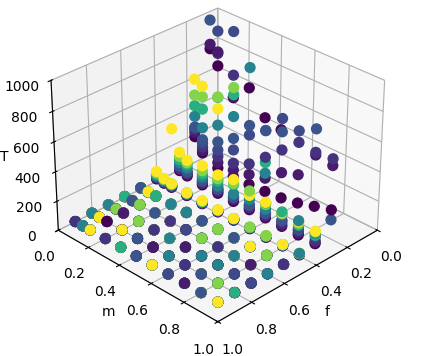}
    \caption{Torus, Moore}
    \label{fig:scatter_tor_mo}
\end{subfigure}
\caption{Mean TTU data for grid-based topologies.}
\label{fig:scatter_grids}
\end{figure}

\begin{figure}
\centering
\begin{subfigure}{.49\linewidth}
    \includegraphics[width=\linewidth]{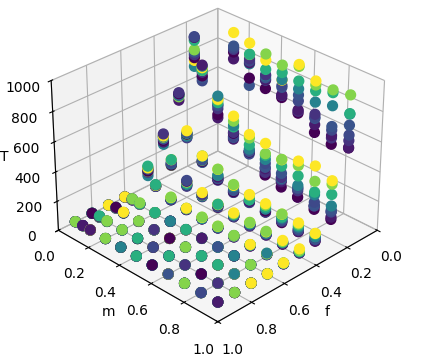}
    \caption{$k=2$}
    \label{fig:scatter_sw_k2}
\end{subfigure}
\begin{subfigure}{.49\linewidth}
    \includegraphics[width=\linewidth]{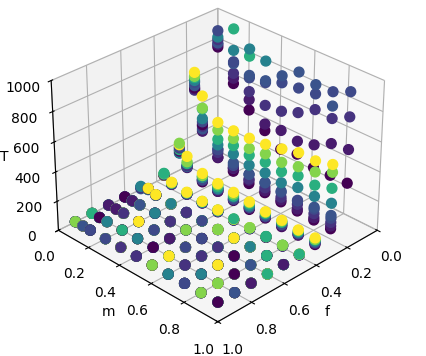}
    \caption{$k=4$}
    \label{fig:scatter_sw_k4}
\end{subfigure}
\begin{subfigure}{.49\linewidth}
    \includegraphics[width=\linewidth]{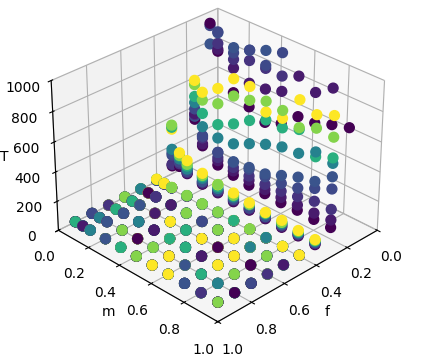}
    \caption{$k=6$}
    \label{fig:scatter_sw_k6}
\end{subfigure}
\begin{subfigure}{.49\linewidth}
    \includegraphics[width=\linewidth]{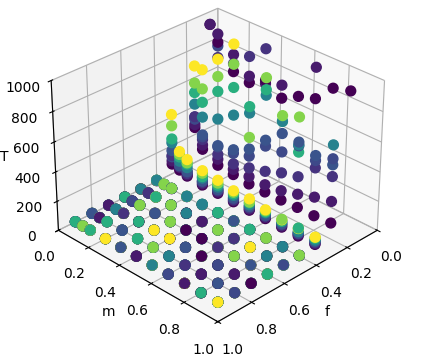}
    \caption{$k=8$}
    \label{fig:scatter_sw_k8}
\end{subfigure}
\caption{Mean TTU data for small-world networks.}
\label{fig:scatter_sw}
\end{figure}

Figures~\ref{fig:scatter_grids} and~\ref{fig:scatter_sw} display TTU means for each topological configuration, ordered by increasing average degree \(d\).

\subsection{Interpretability Notes}

For completeness, we restate the mechanism--term mapping in Equation~\ref{eq:full}:
\begin{itemize*}
    \item Baseline \(1+\cdots\) anchors easy regimes near one step.
    \item \(f^{-p_f}\) enforces divergence as \(f\!\to\!0\) and quantifies seed sensitivity.
    \item \(\ln(M+h_M)\) captures diminishing inertia from memory.
    \item Mobility fraction \(\tfrac{(m+g_m)^{p_m}}{(m+h_m)^{q_m}}\) reproduces rapid early gains and saturation at high \(m\).
    \item Degree factor \(\left(\tfrac{d}{d_{\min}}\right)^{e(f,M)}\) explains why higher degree typically accelerates adoption but can impede it under scarce seeds and large memory; here \(d_{\min}=2-\tfrac{2}{N}\).
\end{itemize*}

These notes bridge the empirical observations of the Results section with the analytical form of Equation~\ref{eq:full}, clarifying how curvature, saturation, and gated interactions correspond to estimated parameters. This alignment supports both quantitative accuracy and mechanistic transparency.





\bibliographystyle{jasss}
\bibliography{references} 


\end{document}